\documentstyle[12pt,aps]{revtex}

\widetext
\draft
\tighten
\begin{document}

\pagestyle{empty}

\thispagestyle{empty}

\title{\normalsize{{\rm{\bf BASE DE HELICIDAD Y PARIDAD}}}\thanks{Ponencia plenaria en la 8a Reuni\'on Nacional Academica de F\'{\i}sica y Matematicas,
12-16 de Mayo del 2003,ESFM-IPN, M\'exico, D. F.}}

\author{{\bf Valeri V. Dvoeglazov}}

\address{{\rm Universidad de Zacatecas, Apartado Postal 636,
Suc. UAZ\\Zacatecas 98062, Zac., M\'exico\\
E-mail: valeri@ahobon.reduaz.mx\\
URL: http://ahobon.reduaz.mx/\~\,valeri/valeri.html}}


\maketitle

\bigskip

\begin{abstract}\hspace*{-10mm} {\bf RESUMEN.} Estudiamos la teor\'{\i}a de
la representaci\'on $(1/2,0)\oplus (0,1/2)$ en la base de helicidad de los espinores correspondientes. 
Como mencionaron Berestetski\u{\i}, Lifshitz y Pitaevski\u{\i}, los autoestados
de helicidad {\it no} son los autoestados de paridad. 
Se discuten relaciones con la teor\'{\i}a cu\'antica de campos del tipo de Gelfand, Tsetlin y Sokolik. Finalmente, se propone una nueva forma del operador de paridad,  el cual conmuta con el Hamiltoniano.
\end{abstract}

\thispagestyle{empty}

\bigskip

\noindent

Recientemente, generalizamos la ecuaci\'on de Dirac~\cite{Barut,DasG,Dv2,Dv2a}
y el formalismo de Bargmann y Wigner\cite{Dv3}. En base a esto propusimos 
un conjunto de 12 ecuaciones para un campo tensorial antisim\'etrico de segundo
rango. Algunas de ellas llevan a transiciones con rompimiento de paridad.
En mi platica voy a presentar un estudio de t\'opicos  que se relacionan con los trabajos anteriores.  ?` C\'omo debe transferirse de base est\'andar a la base de helicidad en la teor\'{\i}a de Dirac?

Es bien conocido que el operador $\hat {\bf S}_3 = {\bbox
\sigma}_3/2\otimes I_2$  no conmuta con el Hamiltoniano de Dirac 
si el 3-momento no est\'a alineado con el tercer eje (usamos expansi\'on 
de la funci\'on de campo en ondas planas):
\begin{equation}
[\hat{\cal H},\hat {\bf S}_3]_- = (\gamma^0{\bbox\gamma}^k \times
{\bf\nabla}_i )_3
\end{equation}
Adem\'as, Berestetski\u{\i}, Lifshitz y
Pitaevsk\u{\i} notaron que~\cite{Lan} ``... el momento orbital angular ${\bf
l}$ y el espin ${\bf s}$ de una part\'{\i}cula en movimiento no se conservan por separado. Unicamente el momento angular total ${\bf j}= {\bf l}+{\bf s}$ se
conserva. Por eso, la componente del esp\'{\i}n en cualquier direcci\'on fija (tomada a lo largo del eje
$z$), tampoco se conserva, y no puede elegirse para enumerar los estados de polarizaci\'on (esp\'{\i}n) de una part\'{\i}cula en movimiento." Lo mismo declar\'o Novozhilov en su libro~\cite{Novozh}.
De otro modo, el operador de helicidad ${\bbox\sigma}\cdot
\widehat {\bf p}/2 \otimes I$, $\widehat {\bf p} = {\bf p}/\vert {\bf
p}\vert$ conmuta con el Hamiltoniano (m\'as precisamente, el conmutador es
igual a cero cuando act\'ua sobre las soluciones de onda plana de una part\'{\i}cula).

A pesar de ello, la gente utiliza la base construida con 4-espinores tales que son auto-estados del operador $\hat {\bf S}_3$:
\begin{eqnarray}
u_{{1\over 2},{1\over 2}} = N_{1\over 2}^+\pmatrix{1\cr0\cr1\cr0\cr}\,,
u_{{1\over 2},-{1\over 2}} =N_{-{1\over 2}}^+ \pmatrix{0\cr1\cr0\cr1\cr}\,,
v_{{1\over 2},{1\over 2}} = N_{1\over 2}^-\pmatrix{1\cr0\cr-1\cr0\cr}\,,
v_{{1\over 2},-{1\over 2}} =N_{-{1\over 2}}^-
\pmatrix{0\cr1\cr0\cr-1\cr}\,.\label{sb1}
\end{eqnarray}
Por otra parte, la base de helicidad 
no ha sido bien estudiada y tampoco es muy usual 
(v\'ease sin embargo~\cite{Novozh,JW,Grei}).  

Los espinores en la base usual son 
\begin{mathletters}
\begin{eqnarray}
u_{{1\over 2},{1\over 2}} = {N_{1\over
2}^+\over \sqrt{2m (E+m)}}
\pmatrix{p^++m\cr p_r\cr p^- +m\cr -p_r\cr}\,,
u_{{1\over 2},-{1\over 2}} ={N_{-{1\over 2}}^+
\over \sqrt{2m (E+m)}}\pmatrix{p_l\cr p^-+m\cr -p_l\cr p^++m\cr}\,, \\
v_{{1\over 2},{1\over 2}} = {N_{1\over
2}^- \over \sqrt{2m (E+m)}}\pmatrix{p^++m\cr p_r\cr -p^- -m\cr p_r\cr}\,,
v_{{1\over 2},-{1\over 2}} ={N_{-{1\over
2}}^- \over \sqrt{2m (E+m)}}\pmatrix{p_l\cr p^-+m\cr p_l\cr -p^+-m\cr}\,,
\end{eqnarray} \end{mathletters}
$p^\pm = E\pm p_z$, $p_{r,l}= p_x\pm ip_y$. Que
son los autoestados de paridad con autovalores $\pm 1$. (Se usa la matr\'{\i}z
$\gamma_0=\pmatrix{0&\openone\cr \openone &
0\cr}$ en el operador de paridad).

Perm\'{\i}tanme atraer su atenci\'on a la base de helicidad.
Los 2-autoespinores del operador de helicidad 
\begin{eqnarray}
{1\over 2} {\bbox \sigma}\cdot\widehat
{\bf p} = {1\over 2} \pmatrix{\cos\theta & \sin\theta e^{-i\phi}\cr
\sin\theta e^{+i\phi} & - \cos\theta\cr}
\end{eqnarray}
pueden parametrizarse de la siguiente forma~\cite{Var,Dv1}:
\begin{eqnarray}
\phi_{{1\over 2}\uparrow}=\pmatrix{\cos{\theta \over 2} e^{-i\phi/2}\cr
\sin{\theta \over 2} e^{+i\phi/2}\cr}\,,\quad
\phi_{{1\over 2}\downarrow}=\pmatrix{\sin{\theta \over 2} e^{-i\phi/2}\cr
-\cos{\theta \over 2} e^{+i\phi/2}\cr}\,,\quad\label{ds}
\end{eqnarray}
para los auto-valores $\pm 1/2$, respectivamente.

Empezamos de la ecuaci\'on de Klein y Gordon generalizada para
descripci\'on de las part\'{\i}culas de esp\'{\i}n 1/2 (i.~e., dos grados de libertad); $c=\hbar=1$:\footnote{El metodo  presentado es gracias a van der Waerden y Sakurai. Recientemente ha sido usado por A. Gersten y V. Dvoeglazov.}
\begin{equation}
(E+{\bbox \sigma}\cdot {\bf p}) (E- {\bbox \sigma}\cdot {\bf p}) \phi
= m^2 \phi\,.\label{de}
\end{equation}
\'Esta puede interpretarse como un conjunto de dos ecuaciones de
primer grado para 2-espinores. Al mismo tiempo, observamos que estos \'ultimos pueden elegirse como auto-estados del operador de helicidad 
que se encuentra, de hecho, en (\ref{de}):\footnote{Esta tesis es contraria
a la elecci\'on de la base (\ref{sb1}), en cual los 4-espinores son autoestados del operador paridad.}
\begin{mathletters}
\begin{eqnarray}
(E-({\bbox\sigma}\cdot {\bf p})) \phi_\uparrow &=& (E-p) \phi_\uparrow
=m\chi_\uparrow \,,\\
(E+({\bbox\sigma}\cdot {\bf p})) \chi_\uparrow &=& (E+p) \chi_\uparrow
=m\phi_\uparrow \,,\\
(E-({\bbox\sigma}\cdot {\bf p})) \phi_\downarrow &=& (E+p) \phi_\downarrow
=m\chi_\downarrow\,, \\
(E+({\bbox\sigma}\cdot {\bf p})) \chi_\downarrow &=& (E-p) \chi_\downarrow
=m\phi_\downarrow \,.
\end{eqnarray}
\end{mathletters}
\normalsize{
Cuando los $\phi$ espinores han sido definidos por la ecuaci\'on (\ref{ds}) podemos construir los correspondientes 4-espinores
$u$ y $v$\,\,\footnote{Se puede tambi\'en
intentar construir otro esquema  diferente de la teor\'{\i}a de Dirac. 
Los 4-espinores podr\'{\i}an ser considerados no como
autoespinores del operador de helicidad en el espacio de representaci\'on $(1/2,0)\oplus (0,1/2)$, cf.~\cite{DasG}. Podr\'{\i}an ser los auto-estados
del operador de helicidad {\it quiral}, que fue  introducido
en~[2a]. En este caso, en lugar de las ecuaciones de Dirac 
las ecuaciones en el espacio de momentos pueden escribirse
(cf.~[2c],\cite{Dv2})
\begin{mathletters}
\begin{eqnarray}
p_\mu \gamma^\mu {\cal U}_\uparrow - m {\cal U}_\downarrow &=&0 \,,\\
p_\mu \gamma^\mu {\cal U}_\downarrow - m {\cal U}_\uparrow &=&0 \,,\\
p_\mu \gamma^\mu {\cal V}_\uparrow + m {\cal V}_\downarrow &=&0 \,,\\
p_\mu \gamma^\mu {\cal V}_\downarrow + m {\cal V}_\uparrow &=&0 \,.
\end{eqnarray}
\end{mathletters}
Las flechas $\uparrow\downarrow$ se refieren a los estados de helicidad quiral, por ejemplo, $u_\eta ={1\over \sqrt{2}} \pmatrix{N \phi_\eta\cr
N^{-1} \phi_{-\eta}\cr}$.}
\begin{mathletters} \begin{eqnarray}
u_\uparrow &=&
N_\uparrow^+ \pmatrix{\phi_\uparrow\cr {E-p\over m}\phi_\uparrow\cr} =
{1\over \sqrt{2}}\pmatrix{\sqrt{{E+p\over m}} \phi_\uparrow\cr
\sqrt{{m\over E+p}} \phi_\uparrow\cr}\,,
u_\downarrow = N_\downarrow^+ \pmatrix{\phi_\downarrow\cr
{E+p\over m}\phi_\downarrow\cr} = {1\over
\sqrt{2}}\pmatrix{\sqrt{{m\over E+p}} \phi_\downarrow\cr \sqrt{{E+p\over
m}} \phi_\downarrow\cr}\,,\label{s1}\\
v_\uparrow &=& N_\uparrow^- \pmatrix{\phi_\uparrow\cr
-{E-p\over m}\phi_\uparrow\cr} = {1\over \sqrt{2}}\pmatrix{\sqrt{{E+p\over
m}} \phi_\uparrow\cr
-\sqrt{{m\over E+p}} \phi_\uparrow\cr}\,,
v_\downarrow = N_\downarrow^- \pmatrix{\phi_\downarrow\cr
-{E+p\over m}\phi_\downarrow\cr} = {1\over
\sqrt{2}}\pmatrix{\sqrt{{m\over E+p}} \phi_\downarrow\cr -\sqrt{{E+p\over
m}} \phi_\downarrow\cr}\,,\label{s2}
\end{eqnarray} \end{mathletters}
donde se ha normalizado  a la unidad
($\pm 1$)):\footnote{Claro, que no existen ningunos dificultades cambiar la normalizaci\'on a $\pm m$, que puede ser m\'as  conveniente
para el estudio del l\'{\i}mite sin masa.}
\begin{mathletters}
\begin{eqnarray}
\bar u_\lambda u_{\lambda^\prime} &=& \delta_{\lambda\lambda^\prime}\,,
\bar v_\lambda v_{\lambda^\prime} = -\delta_{\lambda\lambda^\prime}\,,\\
\bar u_\lambda v_{\lambda^\prime} &=& 0 =
\bar v_\lambda u_{\lambda^\prime}
\end{eqnarray}
\end{mathletters}
Se puede probar que la matr\'{\i}z
\begin{equation}
P=\gamma^0 = \pmatrix{0&\openone\cr\openone & 0\cr}
\label{par}
\end{equation}
es la matr\'{\i}z que se usa en el operador de paridad tanto  en la base
de helicidad como en la base original de la teor\'{\i}a de Dirac. 
De hecho, los 4-espinores
(\ref{s1},\ref{s2}) satisfen la ecuaci\'on de  Dirac en la representaci\'on espinorial
de las matrices  $\gamma$ (v\'ease directamente de (\ref{de})). 
Por eso, la funci\'on transformada bajo paridad
$\Psi^\prime (t, -{\bf x})=P\Psi (t,{\bf x})$ tiene que satisfer
la ecuaci\'on siguiente
\begin{equation}
[i\gamma^\mu \partial_\mu^\prime -m ] \Psi^\prime (t,-{\bf x}) =0 \,,
\end{equation}
con $\partial_\mu^\prime = (\partial/\partial t, -{\bf \nabla}_i)$.
Esto es posible solo cuando $P^{-1}\gamma^0 P = \gamma^0$ y
$P^{-1} \gamma^i P = -\gamma^i$. La matr\'{\i}z (\ref{par})
satisfase los mismos requerimientos, que en los libros de texto.

Sigamos adelante. Es f\'acil mostrar que se pueden formar los operadores de proyecci\'on 
\begin{mathletters}
\begin{eqnarray}
P_+ &=&+\sum_{\lambda} u_\lambda ({\bf p}) \bar u_\lambda ({\bf p})
=\frac{p_\mu \gamma^\mu +m}{2m}\,,\\
P_- &=&-\sum_{\lambda} v_\lambda ({\bf p}) \bar v_\lambda ({\bf p})
= \frac{m- p_\mu \gamma^\mu}{2m}\,,
\end{eqnarray}
\end{mathletters}
con las propiedades conocidas $P_+ +P_- =1$ y $P_\pm^2 = P_\pm$.
Ya conocidos podemos expandir los 4-espinores definidos en la base (\ref{sb1})
como una combinaci\'on lineal de los 4-espinores en la base de helicidad:
\begin{mathletters}
\begin{eqnarray}
u_\sigma ({\bf p}) &=& A_{\sigma\lambda} u_\lambda ({\bf p})
+ B_{\sigma\lambda} v_\lambda ({\bf p})\,,\\
v_\sigma ({\bf p}) &=& C_{\sigma\lambda} u_\lambda ({\bf p})
+ D_{\sigma\lambda} v_\lambda ({\bf p})\,.
\end{eqnarray}
\end{mathletters}
Multiplicando las ecuaciones por $\bar u_{\lambda^\prime}$,
$\bar v_{\lambda^\prime}$ y utilizando las condiciones de normalizaci\'on,
obtenemos $A_{\sigma\lambda}= D_{\sigma\lambda}= \bar u_\lambda u_\sigma$,
$B_{\sigma\lambda}=C_{\sigma\lambda}= - \bar v_\lambda u_\sigma$.
Entonces,  la matr\'{\i}z de transformaci\'on de la base usual a la base de helicidad es
\begin{equation}
\pmatrix{u_\sigma\cr
v_\sigma\cr}={\cal U} \pmatrix{u_\lambda\cr
v_\lambda\cr},\,\,\quad{\cal U} = \pmatrix{A&B\cr
B&A}
\end{equation}
Ni $A$ ni $B$ son unitarias:
\begin{mathletters}
\begin{eqnarray}
A= (a_{++} +a_{+-}) (\sigma_\mu a^\mu) +(-a_{-+} +a_{--})
(\sigma_\mu a^\mu) \sigma_3\,,\\
B= (-a_{++} +a_{+-}) (\sigma_\mu a^\mu) +(a_{-+} +a_{--})
(\sigma_\mu a^\mu) \sigma_3\,,
\end{eqnarray}
\end{mathletters}
donde
\begin{mathletters}
\begin{eqnarray}
a^0 &=& -i\cos (\theta/2) \sin (\phi/2) \in \Im m\,,\quad
a^1 = \sin (\theta/2) \cos (\phi/2)\in \Re e\,,\\
a^2 &=& \sin (\theta/2) \sin (\phi/2) \in \Re e\,,\quad
a^3 = \cos (\theta/2) \cos (\phi/2)\in \Re e\,,
\end{eqnarray}
\end{mathletters}
y
\begin{mathletters}
\begin{eqnarray}
a_{++} &=&\frac{\sqrt{(E+m)(E+p)}}{2\sqrt{2} m}\,,\quad
a_{+-} =\frac{\sqrt{(E+m)(E-p)}}{2\sqrt{2} m}\,,\\
a_{-+} &=&\frac{\sqrt{(E-m)(E+p)}}{2\sqrt{2} m}\,,\quad
a_{--} =\frac{\sqrt{(E-m)(E-p)}}{2\sqrt{2} m}\,.
\end{eqnarray}
\end{mathletters}
Sin embargo, $A^\dagger A+B^\dagger B =\openone$, que lleva a la conclusi\'on de que la matr\'{\i}z de $4\times 4$ ${\cal U}$
es unitaria. Esta matr\'{\i}z act\'ua
a los \'{\i}ndices de  {\it esp\'{\i}n} ($\sigma$,$\lambda$), {\it no} a
los indices espinoriales. La transformaci\'on tambi\'en 
se presenta en la siguiente forma:
\begin{mathletters}
\begin{eqnarray}
u_\sigma^\alpha &=& [A_{\sigma\lambda}\otimes I_{\alpha\beta}
+B_{\sigma\lambda}\otimes \gamma^5_{\alpha\beta}] u_\lambda^\beta\,,\\
v_\sigma^\alpha &=& [A_{\sigma\lambda}\otimes I_{\alpha\beta}
+B_{\sigma\lambda}\otimes \gamma^5_{\alpha\beta}] v_\lambda^\beta\,.
\end{eqnarray}
\end{mathletters}

Vamos a investigar las propiedades de los 4-espinores en base de helicidad
con respecto a las operaciones de simetrias discretas $P,C$ y $T$.
Se espera que $\lambda\rightarrow -\lambda$ con respecto a la paridad $P$,
como dicen Berestetski\u{\i}, Lifshitz y Pitaevski\u{\i}~\cite{Lan}.\footnote{De hecho, 
si ${\bf x}\rightarrow -{\bf x}$,
entonces el vector ${\bf p}\rightarrow -{\bf p}$, pero el pseudovector 
${\bf S}\rightarrow {\bf S}$, que implica la declaraci\'on arriba.}
Para ${\bf p}$ tenemos ${\bf p} \rightarrow -{\bf p}$ (lo que implica en un sistema de coordenadas esf\'ericas que los \'angulos cambian seg\'un
$\theta \rightarrow \pi-\theta$,
$\varphi \rightarrow \pi+\varphi$) y los 2-autoespinores de helicidad
transforman en la siguiente manera: $\phi_{\uparrow\downarrow} \Rightarrow
-i \phi_{\downarrow\uparrow}$, v\'ease~\cite{Dv1}.
Entonces,
\begin{mathletters}
\begin{eqnarray}
Pu_\uparrow (-{\bf p}) &=& -i u_\downarrow ({\bf p})\,,
Pv_\uparrow (-{\bf p}) = +i v_\downarrow ({\bf p})\,,\\
Pu_\downarrow (-{\bf p}) &=& -i u_\uparrow ({\bf p})\,,
Pv_\downarrow (-{\bf p}) = +i v_\uparrow ({\bf p})\,.
\end{eqnarray}
\end{mathletters}
Concluimos que: en un nivel cl\'asico, observamos que
los 4-espinores de helicidad transforman a los 4-espinores de helicidad opuesta.

Con respecto a la operaci\'on  de conjugaci\'on de carga:
\begin{equation}
C =\tilde C {\cal K} =\pmatrix{0&\Theta\cr
-\Theta & 0\cr} {\cal K}
\end{equation}
tenemos
\begin{mathletters}
\begin{eqnarray}
Cu_\uparrow ({\bf p}) &=& - v_\downarrow ({\bf p})\,,
Cv_\uparrow ({\bf p}) = +  u_\downarrow ({\bf p})\,,\\
Cu_\downarrow ({\bf p}) &=& + v_\uparrow ({\bf p})\,,
Cv_\downarrow ({\bf p}) = - u_\uparrow ({\bf p})\,,
\end{eqnarray}
\end{mathletters}
gracias a las propiedades del operador de Wigner $\Theta \phi_\uparrow^\ast =
-\phi_\downarrow$ y $\Theta \phi_\downarrow^\ast = \phi_\uparrow$.
En el caso de la operaci\'on $CP$ (y $PC$) obtenemos:
\begin{mathletters}
\begin{eqnarray}
C P u_\uparrow (-{\bf p}) &=& -PC u_\uparrow (-{\bf p})
= +i v_\uparrow ({\bf p})\,,\\
C P u_\downarrow
(- {\bf p}) &=& - P C u_\downarrow (-{\bf p}) = -i v_\downarrow ({\bf
p})\,,\\
C P v_\uparrow (-{\bf p}) &=& - P C v_\uparrow (-{\bf p}) =
+  i u_\uparrow ({\bf p})\,,\\
C P v_\downarrow (-{\bf p}) &=& - P C v_\downarrow (-{\bf p}) =
- i u_\downarrow ({\bf p})\,.
\end{eqnarray}
\end{mathletters}
Conclusiones similares tambi\'en pueden ser figuradas en el espacio de Fock. 
Definimos el operador de campo como:
\begin{equation}
\Psi (x^\mu) = \sum_\lambda \int \frac{d^3 {\bf p}}{(2\pi)^3}
\frac{\sqrt{m}}{2E} [ u_\lambda a_\lambda e^{-ip_\mu x^\mu} +v_\lambda
b^\dagger_\lambda e^{+ip_\mu x^\mu} ]\,.
\end{equation}
Se asume que las relaciones de conmutaci\'on son las relaciones est\'andar~\cite{Bogol,Wein,Itzyk,Greib}\footnote{Los unicos cambios posibles
pueden ser relacionados con  formas diferentes de normalizaci\'on de los 4-espinores, los cuales contribuyen un factor adicional de la funci\'on $\delta$.}
(compare con~\cite{DasG,Dv2})
\begin{mathletters}
\begin{eqnarray}
\left [a_\lambda ({\bf p}),
a_{\lambda^\prime}^\dagger ({\bf k})
\right ]_+ &=& 2E \delta^{(3)} ({\bf p}-{\bf k})
\delta_{\lambda\lambda^\prime}\,,
\left [a_\lambda ({\bf p}),
a_{\lambda^\prime} ({\bf k})\right ]_+ = 0 =
\left [a_\lambda^\dagger ({\bf p}),
a_{\lambda^\prime}^\dagger ({\bf k})\right ]_+\,,\\
\left [a_\lambda ({\bf p}),
b_{\lambda^\prime}^\dagger ({\bf k})\right ]_+ &=& 0 =
\left [b_\lambda ({\bf p}),
a_{\lambda^\prime}^\dagger ({\bf k})\right ]_+\,,\\
\left [b_\lambda ({\bf p}),
b_{\lambda^\prime}^\dagger ({\bf k})
\right ]_+ &=& 2E \delta^{(3)} ({\bf p}-{\bf k})
\delta_{\lambda\lambda^\prime}\,,
\left [b_\lambda ({\bf p}),
b_{\lambda^\prime} ({\bf k})\right ]_+ = 0 =
\left [b_\lambda^\dagger ({\bf p}),
b_{\lambda^\prime}^\dagger ({\bf k})\right ]_+\,.
\end{eqnarray} \end{mathletters}
\normalsize{
Si uno define $U_P \Psi (x^\mu) U_P^{-1} = \gamma^0  \Psi
(x^{\mu^\prime})$, $U_C \Psi (x^\mu) U_C^{-1} = C \Psi^\dagger (x^\mu)$
y el operador anti-unitario de reversi\'on de tiempo $(V_T \Psi (x^\mu)
V_T^{-1})^\dagger = T \Psi^\dagger (x^{\mu^{\prime\prime}})$,
entonces es f\'acil obtener las transformaciones corespondientes
de los operadores de creaci\'on y anniquilaci\'on (cf.  los libros de texto citados).
\begin{mathletters}
\begin{eqnarray}
U_P a_\lambda U_P^{-1} &=& -i a_{-\lambda} (-{\bf p})\,,
U_P b_\lambda U_P^{-1} = -i b_{-\lambda} (-{\bf p})\,,\label{pa1}\\
U_C a_\lambda U_C^{-1} &=& (-1)^{{1\over 2}+\lambda} b_{-\lambda} ({\bf
p})\,, U_C b_\lambda U_C^{-1} = (-1)^{{1\over 2}-\lambda} a_{-\lambda}
(-{\bf p})\,.
\end{eqnarray}
\end{mathletters}
Como consequencia, obtenemos (si el vac\'{\i}o tiene paridades postivas $U_P \vert 0>=\vert 0>$,
$U_C\vert 0>= \vert 0>$)
\begin{mathletters}\begin{eqnarray}
&&U_P a^\dagger_\lambda ({\bf p}) \vert 0>=
U_P a_\lambda^\dagger U_P^{-1} \vert 0> =i a_{-\lambda}^\dagger
(-{\bf p}) \vert 0>= i \vert -{\bf p}, -\lambda >^+\,,\\ &&U_P
b^\dagger_\lambda ({\bf p}) \vert 0> =U_P b_\lambda^\dagger U_P^{-1}
\vert 0> =i b_{-\lambda}^\dagger (-{\bf p}) \vert 0>= i \vert -{\bf p},
-\lambda >^-\,;
\end{eqnarray}\end{mathletters}
y
\begin{mathletters}\begin{eqnarray}
&&U_C a^\dagger_\lambda ({\bf p}) \vert 0>=
U_C a_\lambda^\dagger U_C^{-1} \vert 0> = (-1)^{{1\over 2} +\lambda}
b_{-\lambda}^\dagger ({\bf p}) \vert 0>= (-1)^{{1\over 2}+\lambda} \vert
{\bf p}, -\lambda >^-\,,\\
&&U_C b^\dagger_\lambda ({\bf p}) \vert 0>= U_C
b_\lambda^\dagger U_C^{-1} \vert 0> = (-1)^{{1\over
2}-\lambda} a_{-\lambda}^\dagger ({\bf p}) \vert 0>= (-1)^{{1\over
2}-\lambda} \vert {\bf p}, -\lambda >^+\,.
\end{eqnarray}\end{mathletters}
Finalmente, para la operaci\'on $CP$ obtenemos:
\begin{mathletters}
\begin{eqnarray}
&&U_P U_C a^\dagger_\lambda ({\bf p}) \vert 0>=
-U_C U_P a^\dagger_\lambda ({\bf p}) \vert 0> = (-1)^{{1\over 2}+\lambda}
U_P b_{-\lambda}^\dagger \vert 0> =\nonumber\\
&=& i (-1)^{{1\over 2} +\lambda}
b_{\lambda}^\dagger (-{\bf p}) \vert 0>= i(-1)^{{1\over 2}+\lambda} \vert
-{\bf p}, \lambda >^-\,,\\
&&U_P U_C b^\dagger_\lambda ({\bf p}) \vert 0>= -U_C U_P b^\dagger_\lambda
({\bf p}) = (-1)^{{1\over 2}-\lambda} U_P
a_{-\lambda}^\dagger \vert 0> = \nonumber\\
&=&i (-1)^{{1\over
2}-\lambda} a_{\lambda}^\dagger (-{\bf p}) \vert 0>= i(-1)^{{1\over
2}-\lambda} \vert -{\bf p}, \lambda >^+\,,
\end{eqnarray}
\end{mathletters}
\normalsize{
lo que significa que las operaciones $P$ y $C$ anticonmutan en el caso
de la representaci\'on $({1\over 2},0)\oplus (0,{1\over 2})$, lo cu\'al
es opuesto
a la teor\'{\i}a basada en  4-espinores, auto-estados de  helicidad quiral
(cf.~\cite{Dv2}).

Debido a que $V_T$ es un operador  anti-unitario, hay que tomar cuenta que en este caso 
los $c$-numeros se ponen afuera de operaci\'on de conjugaci\'on hermitica
{\it sin} conjugaci\'on compleja:
\begin{equation}
[V_T \lambda A V_T^{-1}]^\dagger = [\lambda^\ast V_T A V_T^{-1} ]^\dagger
= \lambda [V_T A^\dagger V_T^{-1} ]\,.
\end{equation}
Tomando en cuenta esta definici\'on obtenemos:\footnote{Se escoge $T$, 
$T=\pmatrix{\Theta &0\cr
0& \Theta\cr}$ de manera que se satisfagan las condiciones
$T^{-1} \gamma_0^T T= \gamma_0$, $T^{-1} \gamma_i^T T= \gamma_i$
y $T^T= -T$.}
\begin{mathletters}
\begin{eqnarray}
V_T a_\lambda^\dagger V_T^{-1} &=& +i
(-1)^{{1\over 2}-\lambda}
a_{\lambda}^\dagger (-{\bf p})\,,\\
V_T b_\lambda  V_T^{-1} &=& +i (-1)^{{1\over 2}-\lambda}
b_{\lambda} (-{\bf p})\,.
\end{eqnarray}
\end{mathletters}

Adem\'as, observamos que la respuesta a la pregunta de si
una part\'{\i}cula y su antipart\'{\i}cula tienen
paridades iguales u opuestas paridades, depende de un factor de fase seg\'un  la siguiente
expresi\'on:
\begin{equation}
U_P \Psi (t, {\bf x}) U_P^{-1} = e^{i\alpha} \gamma^0 \Psi (t, -{\bf
x})\,.  \label{def1}
\end{equation}
De hecho, si repetimos el procedimiento de los libros de texto~\cite{Greib},
tenemos:
\begin{eqnarray}
\lefteqn{U_P \left [ \sum_{\lambda}^{}
\int {d^3 {\bf p} \over (2\pi)^3} {\sqrt{m}\over 2E}
(u_\lambda ({\bf p}) a_\lambda ({\bf p}) e^{-ip_\mu x^\mu}
+v_\lambda ({\bf p}) b_\lambda^\dagger ({\bf p}) e^{+ip_\mu x^\mu}) \right
] U_P^{-1} =\,\nonumber}\\
&=& e^{i\alpha} \left [ \sum_{\lambda}^{} \int {d^3 {\bf
p} \over (2\pi)^3} {\sqrt{m}\over 2E} (\gamma^0 u_\lambda (-{\bf
p}) a_\lambda (-{\bf p}) e^{-ip_\mu x^\mu} +\gamma^0 v_\lambda
(-{\bf p}) b_\lambda^\dagger (-{\bf p}) e^{+ip_\mu x^\mu}) \right ]= \,
\nonumber\\
&=& e^{i\alpha} \left [ \sum_{\lambda}^{} \int {d^3 {\bf
p} \over (2\pi)^3} {\sqrt{m}\over 2E} (-i u_{-\lambda} ({\bf
p}) a_\lambda (-{\bf p}) e^{-ip_\mu x^\mu} +i v_\lambda
({\bf p}) b_\lambda^\dagger (-{\bf p}) e^{+ip_\mu x^\mu}) \right ]\,.
\end{eqnarray}
Multiplicando por $u_{\lambda^\prime} ({\bf p})$
y $v_{\lambda^\prime} ({\bf p})$ consecutivamente,
y usando las condiciones de normalizaci\'on, obtenemos
\begin{mathletters}
\begin{eqnarray}
&&U_P a_\lambda U_P^{-1} = -i e^{i\alpha} a_{-\lambda} (-{\bf p})\,,\\
&&U_P b_\lambda^\dagger U_P^{-1} = + i e^{i\alpha} b_{-\lambda}^\dagger
(-{\bf p})\,.
\end{eqnarray}
\end{mathletters}
Como se ve, si $\alpha=\pi/2$ obtenemos
las paridades {\it opuestas} de los operadores de  creaci\'on y aniquilaci\'on
para part\'{\i}culas y anti-part\'{\i}culas:
\begin{mathletters}
\begin{eqnarray}
&&U_P a_\lambda U_P^{-1} = + a_{-\lambda} (-{\bf p})\,,\\
&&U_P b_\lambda U_P^{-1} = - b_{-\lambda}
(-{\bf p})\,.
\end{eqnarray}
\end{mathletters}
Sin embargo, diferencia con el caso de Dirac sigue existiendo
($\lambda$ transforma a $-\lambda$). Como conclusi\'on, vemos que
la cuesti\'on de paridades relativas intr\'{\i}nsecas iguales (opuestas)
est\'a relacionada con el factor de fase en la ecuaci\'on (\ref{def1}).
Recu\'erdese que en cierto sentido tenemos una situaci\'on parecida en la construcci\'on  del operador de campo para neutrinos (cf. con el factor de fase de Goldhaber y Kayser).

Finalmente, busquemos la forma expl\'{\i}cita del operador de paridad $U_P$.
Probaremos que el conmuta con el Hamiltoniano.
Vamos utilizar el m\'etodo presentado en~\cite[\S 10.2-10.3]{Greib}.
\'Este est\'a basado en el {\it anzatz} de que $U_P = \exp [i\alpha \hat A] \exp [i \hat
B]$ con $\hat A =\sum_s\int d^3 {\bf p} [a_{{\bf p},s}^\dagger a_{-{\bf
p}s} +b_{{\bf p}s}^\dagger b_{-{\bf p}s}]$ y $\hat B =\sum_s\int d^3
{\bf p} [\beta a_{{\bf p},s}^\dagger a_{{\bf p}s} +\gamma b_{{\bf
p}s}^\dagger b_{{\bf p}s}]$. Utilizando la identidad
\begin{equation}
e^{\hat A} \hat B e^{-\hat A} = \hat B +[\hat A,\hat B]_- +{1\over 2!}
[\hat A, [\hat A,\hat B]]+\ldots
\end{equation}
y $[\hat A,\hat B\hat C]_-= [\hat A,\hat B]_+ \hat C
-\hat B [\hat A,\hat C]_+$ se puede fijar los parametros
$\alpha,\beta,\gamma$ de tal manera que se satisfaga el requerimiento f\'{\i}sico de que las part\'{\i}culas de Dirac tengan paridades opuestas.

En nuestro caso, necesitamos satisfacer (\ref{pa1}), i.e., el operador
tiene que invertir no solo el signo del momento, sino tambi\'en el signo de la helicidad. Esto puede hacerse  proponiendo un postulado parecido:
$U_P= e^{i\alpha \hat A}$ con
\begin{equation}
\hat A =\sum_{s}^{} \int {d^3 {\bf p}\over 2E}
[a^\dagger_\lambda ({\bf p}) a_{-\lambda} (-{\bf p})
+b_\lambda^\dagger ({\bf p}) b_{-\lambda} (-{\bf p})]\,.
\end{equation}
Como puede comprobarse directamente, las ecuaciones (\ref{pa1})
se satisfacen si $\alpha=\pi/2$. Hay que comparar  este operador de paridad con el  que fue dado en~\cite{Itzyk,Greib} para el campo de Dirac:\footnote{Greiner utiliz\'o las siguientes relaciones de conmutaci\'on
$\left [ a ({\bf p}, s), a^\dagger ({\bf p}^\prime, s^\prime) \right ]_+ =
\left [ b ({\bf p}, s), b^\dagger ({\bf p}^\prime, s^\prime) \right ]_+ =
\delta^3 ({\bf p}-{\bf p}^\prime) \delta_{ss^\prime}$. Se debe notar que 
la forma de Greiner del operador de paridad no es \'unica.
Itzykson y Zuber~\cite{Itzyk} propusieron otra, que difiere  
de (10.69) del trabajo~\cite{Greib} por varios factores de fase.
Para buscar las relaciones entre estas dos formas
de operadores de paridad se debe aplicar una rotaci\'on adicional 
en el espacio de Fock.}
\begin{eqnarray} \lefteqn{U_P =
\exp \left [ i{\pi\over 2} \int  d^3 {\bf p} \sum_s
\left ( a ({\bf p}, s)^\dagger
a (\tilde{\bf p},s) +b ({\bf p},s)^\dagger b (\tilde{\bf p},s)-
\right.\right.}\nonumber\\
&&\left.\left.- a ({\bf p},s)^\dagger a ({\bf p},s) + d ({\bf
p},s)^\dagger b ({\bf p},s) \right ) \right ]\,,\quad (10.69)\,\,
\mbox{del}\,\,\,
\mbox{trabajo}~\cite{Greib}\end{eqnarray}
Verificando directamente se puede concluir que nuestro nuevo operador $U_P$ conmuta
con el Hamiltoniano:  \begin{equation} {\cal H} = \int d^3 {\bf x}
\Theta^{00} = \int d^3 {\bf k} \sum_\lambda [ a_\lambda^\dagger ({\bf k})
a_\lambda ({\bf k}) - b_\lambda ({\bf k}) b_\lambda^\dagger ({\bf k})]\,,
\end{equation} i.e.
\begin{equation}
[U_P, {\cal H} ]_- =0\,.
\end{equation}
Por cierto, podemos intentar escoger otros conjuntos adem\'as de las relaciones de conmutaci\'on~[2b,3] (por ejemplo, considerando caso del conjunto de estados
bi-ortonormales), lo que dejo para las publicaciones futuras.

Finalmente, ya que mis trabajos 
recientes est\'an relacionados con lo que se llama ``la teor\'{\i}a
cu\'antica de campos del tipo de 
Bargmann, Wightman y Wigner", quisiera tomar la oportunidad para se\~nalar algunos errores en la interpretaci\'on de varios autores.
Este tipo de teor\'{\i}as ha sido propuesto por primera vez por Gel'fand y
Tsetlin~[16a]; de hecho, est\'an basadas en la representaci\'on del grupo de inversi\'on de dimensi\'on 2. Es aplicable cuando uno quiere construir una teor\'{\i}a en la que  $C$ y $P$ anticonmutan. Gel'fand, Tsetlin y Sokolik indicaron la posibilidad de aplicarla a la descripci\'on del conjunto
de $K$-mesones y especularon sobre relaciones con el resultado de Lee e Yang.
La conmutatividad/anticonmutatividad de las operaciones de simetr\'{\i}as discretas  
ha sido investigada, tambi\'en, por
Foldy y Nigam~\cite{FN}. Se han discutido posibles relaciones  en [16b] y en los trabajos posteriores de Sokolik
de la construcci\'on de
Gel'fand y Tsetlin con la representaci\'ones del grupo de
anti-deSitter $SO(3,2)$ y con la teor\'{\i}a de la relatividad general
(incluyendo las transformaciones  continuas y discretas). E. Wigner~\cite{Wig}
present\'o los resultados en la Escuela de Istanbul de F\'{\i}sica Te\'orica
en 1962. M\'as tarde, Fushchich discuti\'o
las correspopndientes ecuaciones de onda. Finalmente, en el trabajo~\cite{Dva}
los autores se refirieron a la teor\'{\i}a, en la que un bos\'on y su anti-bos\'on
tienen paridades intr\'{\i}nsecas opuestas como en ``la teor\'{\i}a del tipo de 
Bargmann-Wightman-Wigner". Actualmente, dicha teor\'{\i}a representa 
una generalizaci\'on de tipo de Dirac (!) de la teor\'{\i}a de Weinberg con $2(2J+1)$
componentes para el esp\'{\i}n 1. Se ha expuesta anteriormente por Sankaranarayanan
y Good en el trabajo de  1965, ref.~\cite{SG}. En el art\'{\i}culo~[19b] (y en las
preimpresiones antecedentes del IF-UNAM de 1994) yo present\'e una teor\'{\i}a
basada en un conjunto  de ecuaciones del tipo de Weinberg, cada uno 
para la funci\'on de campo con 6 componentes (que he nombrado los como
``dubletos de Weinberg"). Se ha propuesto en Ref.~[2b,2c,3] la teor\'{\i}a en la representaci\'on $({1\over
2},0)\oplus (0,{1\over 2})$, basada en los 4-autoespinores de helicidad quiral. Se han descubierto relaciones  con la consideraci\'on de Foldy
y Nigam. Se han obtenido las ecuaciones correspondientes
en~\cite{Dv2} y otros art\'{\i}culos.
Sin embargo, posteriormente encontramos art\'{\i}culos de Ziino y Barut~\cite{Barut}
entre otros, que tambi\'en tienen relaciones con la cuesti\'on en discusi\'on.
La teor\'{\i}a similar se construye si definimos los operadores de campos como:
\begin{mathletters}
\begin{eqnarray}
\Psi_1 &=&
\int {d^3 {\bf p}\over (2\pi)^3}
{\sqrt{m}\over 2E} \left [(u_\uparrow a_\uparrow
+v_\uparrow b_\uparrow) e^{-ip_\mu x^\mu}
+ (u_\uparrow a_\uparrow^\dagger
+v_\uparrow b_\uparrow^\dagger) e^{+ip_\mu x^\mu}\right ]\,,\\
\Psi_2 &=&
\int {d^3 {\bf p}\over (2\pi)^3}
{\sqrt{m}\over 2E} \left [(u_\downarrow a_\downarrow
-v_\downarrow b_\downarrow) e^{-ip_\mu x^\mu}
+ (u_\downarrow a_\downarrow^\dagger
-v_\downarrow b_\downarrow^\dagger) e^{+ip_\mu x^\mu}\right ]\,.
\end{eqnarray}
\end{mathletters}

Conclusiones de mi platica:

\begin{itemize}

\item
Similarmente al caso de la representaci\'on $({1\over 2},{1\over 2})$,
las funciones de campo de la representaci\'on $({1\over 2},0)\oplus (0,{1\over 2})$ en base de helicidad 
{\it no} son autoestados del operador paridad; $\vert {\bf p},\lambda> \Rightarrow
\vert -{\bf p},-\lambda >$ tanto a nivel cl\'asico como a nivel cu\'antico. 
Esto est\'a de acuerdo con la consideraci\'on anterior de Berestetski\u{\i},
Lifshitz y Pitaevski\u{\i}.

\item
Las funciones de campo en la base de helicidad satisfacen la ecuaci\'on ordinaria de Dirac, estando las 
$\gamma$'s en la representaci\'on espinorial. Mientras tanto, las funciones de campo en la base de helicidad quiral satisfacen las ecuaciones de la forma
$\hat p \Psi_1 - m\Psi_2 =0$.

\item
Las funciones de campo en la base de helicidad pueden ser expandidas en los 4-espinores ordinarios por medio de la matr\'{\i}z ${\cal U}^{-1}$ que esta dada en el art\'{\i}culo.
Ni $A$, ni $B$ son unitarias, sin embargo, $A^\dagger A+B^\dagger B =
\openone$.

\item
Las operaciones $P$ y $C$ anticonmutan  en este contexto, tanto a nivel cl\'asico como a nivel cu\'antico (esto es opuesto a la teor\'{\i}a basada en los autoestados de helicidad quiral~\cite{Dv2}).

\item
Las part\'{\i}culas y antipart\'{\i}culas pueden ser de paridad igual u opuesta 
con respecto del operador correspondiente. El resultado depende de la elecci\'on del factor de fase en
$U_P \Psi U_P^{-1} = e^{i\alpha} \gamma^0 \Psi^\prime$; por otra parte, se obtiene lo mismo (cambio de las propiedades) por medio  de una rotaci\'on adicional $U_{P_2}$.

\item
Las confusiones antecedentes respecto  de la teor\'{\i}a cu\'antica de campos del tipo  de Gelfand, Tsetlin y Sokolik, Nigam y Foldy, Bargmann, Wightman y Wigner
(GTsS-NF-BWW) han sido aclaradas.

\end{itemize}

Estoy agradecido al Prof. Z. Oziewicz y los participantes de
los Congresos de Algebras de Clifford y ``Graph-Operads-Logic"
(Mayo del 2002).  Reconozco la ayuda en la ortograf\'{\i}a
espa\~nola del Sr. J. L. Quintanar Gonz\'alez.

Universidad de Zacatecas, M\'exico, se agradece por la concesi\'on
del profesorado.  Este trabajo ha sido apoyado en parte por
el Sistema Nacional de Investigadores y el Programa de Apoyo
a la Carrera Docente.

\nopagebreak

\end{document}